\documentclass[10pt]{iopart}
\usepackage{iopams}
\usepackage{setstack}
\usepackage{graphicx}

\newcommand{\lscat}{\ell_{\rm s}}
\newcommand{\ltrans}{\ell_{\rm t}}

\begin{document}

\title{Weak disorder corrections of the scattering and transport mean free path}

\author{Felix Eckert, Andreas Buchleitner and Thomas Wellens}

\address{Physikalisches Institut, Albert-Ludwigs-Universit\"at Freiburg, Hermann-Herder-Str. 3, 79104 Freiburg, D}
\eads{\mailto{felix.eckert@pluto.uni-freiburg.de}, \mailto{andreas.buchleitner@physik.uni-freiburg.de}, \mailto{thomas.wellens@physik.uni-freiburg.de}}

\begin{abstract}
We present a calculation of the weak localisation correction to the scattering and to the transport mean free path, for waves propagating in a $\delta$-correlated random potential, going beyond the usual diffusion approximation for the loops in the interfering path amplitudes. We find that not only the transport mean free path, but also the scattering mean free path is affected by the interference contributions. We also find the dependence of the leading order contribution to both, the scattering and the transport mean free path, on the disorder parameter $1/k\ell$ to be linear. This is in contrast to the result obtained from the diffusion approximation according to which the scattering mean free path remains unaffected when changing the disorder parameter, whereas the correction of the transport mean free path scales like $1/(k\ell)^{2}$.
\end{abstract}

\submitto{\JPA}


\section{Introduction}\label{s:introduction}
A thorough understanding of the propagation of waves in disordered potentials is of great importance in many areas of physics. Classically, one expects diffusive transport of the average wave intensity through the medium, i.e.\ one expects that interference effects play no role, on average. In other words, the neglect of interference results in the average wave intensity performing a random walk just like a classical particle would do. This picture was applied quite successfully to the propagation of electrons in conductors leading to the ohmic dependence of the transmission on the sample thickness \cite{akkermans:mesoscopic} and, in optics, to the theory of radiative transfer \cite{chandrasekhar:radiativetransfer}. 

In general, diffusive transport is characterized by the fact that the square of the distance $r$ of the particle from its initial position increases, on average, linearly with time, i.e., $\langle r^2\rangle = 2 d D t$, with diffusion constant $D$, in $d$ dimensions. We will here consider the case $d=3$. If particles are continuously emitted from a source, the particle density evolves into a \emph{stationary} distribution, which, in the diffusive case, decays asymptotically like $1/r$, with a prefactor which is solely determined by the strength of the source and the {\em transport mean free path} $\ltrans$ (see \sref{sec:diffusion} for more details). The transport mean free path thus characterizes the behaviour of the stationary particle density at \emph{large} distances $r$ from the source, or -- vice versa -- the latter can be used to {\em define} the former. On the other hand, the {\em scattering mean free path} $\lscat$, i.e., the average distance between two successive scattering events, determines the density of particles at \emph{small} distances, where the particles have not yet been scattered after emission from the source (for waves, this particle density corresponds to the \emph{coherent intensity}). This contribution to the density of particles decays exponentially on the length scale $\lscat$. It is thus possible to measure the scattering mean free path via detection of the coherent intensity as a function of the sample thickness. For isotropic scattering of classical particles, transport and scattering mean free path are identical, whereas, in the general case, they are related by $\ltrans = \lscat/(1-\langle \cos(\theta) \rangle)$, where $\langle \cos(\theta) \rangle$ is the angular average over the cosine of the scattering angle. The transport mean free path therefore defines the length scale over which the particles' direction of propagation is randomised. For charged particles such as electrons, another important quantity describing stationary transport is the electrical \emph{conductivity} $\sigma$, defined by the diffusive current in response to an externally applied electric field. In the classical case, the conductivity is proportional to the transport mean free path, i.e., $\sigma=ne^2\ltrans/mv$, called \emph{Drude conductivity}, with electronic density $n$, charge $e$, mass $m$ and transport velocity $v$ (see, e.g., chapter 7.2 of \cite{akkermans:mesoscopic}). Finally, also the diffusion constant is related to $\ltrans$ via $D=\ltrans v/3$.

Coming back to the transport of waves, the above considerations referring to the random walk of a classical particle neglect one important feature inherent to waves -- their ability for interference. The Boltzmann approximation -- which yields the above classical results, and which is valid for wave lengths much smaller than the scattering mean free path -- assumes that, upon averaging over the disorder, all interference contributions vanish. However, if the system exhibits \emph{reciprocity symmetry} \cite{tiggelen:1997} certain interference contributions {\em do} survive the disorder average and lead to deviations from the classical results. The most important contribution originates from the so called \emph{maximally crossed} scattering paths \cite{watson:688,langerneal:prl16}, where the two amplitudes $\psi$ and $\psi^*$ forming the wave intensity $|\psi|^2$ visit the same scatterers, but in opposite order. In a backscattering experiment, these amplitudes will interfere constructively in retro reflection giving rise to the famous \emph{coherent backscattering} effect \cite{albada:prl55}. Analogously, this effect enhances the probability for the wave energy to return to any point inside the medium and thus, because of flux conservation, reduces the overall transport over large distances -- an effect known as {\em weak localisation} \cite{bergmann:physrep107}. Nevertheless -- in three dimensions and if the disorder is not too strong (i.e. if the scattering mean free path is still larger than the wavelength) -- the propagation of the average intensity still resembles a diffusion process, although the two parameters $\lscat$ and $\ltrans$ -- which, as discussed above, govern the behaviour of the stationary intensity distribution on small and large length scales, respectively -- are then modified with respect to their classical values. In particular, the weak localisation effect manifests itself as a reduction of the transport mean free path.

In most of the standard literature, e.g.\ \cite{akkermans:mesoscopic,rammer:qtt,sheng:2006}, the weak localisation correction is calculated for the diffusion constant instead of the transport mean free path, and, furthermore, a cut-off procedure is usually employed which relies on the following argument: as described above, weak localisation is related to the probability for the wave intensity to return to a position where it has been previously. To obtain a finite quantity, one considers a finite small volume around the initial position to determine the return probability. In momentum space, this amounts to a cut-off at large momenta the precise value of which depends on the model under consideration \cite{tiggelen:epl15,tiggelen:2001}. Furthermore, the return probability is usually evaluated within the diffusion approximation, assuming that many scattering events occur before the wave intensity returns to its previous position. As a result of these approximations, the reduction of the diffusion constant turns out to scale like $1/(k\ell)^2$, where $k$ denotes the wave number.

In the present paper, we relax these approximations and calculate the corrections of the scattering and of the transport mean free path without employing any assumption apart from $k\ell\gg 1$ (\lq weak disorder\rq). For simplicity, we will concentrate on the case of a white-noise disorder potential, where, in the Boltzmann approximation, scattering is isotropic, and, hence, the classical values for the scattering and the transport mean free path are identical: $\ltrans^{(\rm cl)}=\lscat^{(\rm cl)}=\ell$. We will then determine the leading modifications of $\lscat$ and $\ltrans$ due to interference effects for $k\ell \gg 1$. For both quantities, the leading corrections turn out to scale like $1/(k\ell)$. Let us note that a similar calculation of the leading weak disorder corrections has already been performed for the conductivity $\sigma$ (which is defined by the response to an external field, e.g.\ in the case of electronic transport) \cite{kirkpatrick:prb34,wysokinski:pre52}. In the present paper, we repeat this analysis for the scattering and transport mean free path, which -- as explained above -- are general quantities determining the stationary intensity distribution for any kind of wave undergoing a diffusive scattering process in the absence of an external field (which is the typical situation, e.g.\ for light waves or sound waves). As we will see, the weak disorder correction of $\sigma$ is identical to the correction of the product of the scattering and of the transport mean free path.

\medskip
\noindent
The paper is structured as follows:
\begin{itemize}
	\item The \emph{next section} presents a short review of the diagrammatic calculation of the average wave intensity. For this, we first introduce the model under investigation and derive the vacuum Green's function. Then, the \emph{disorder average} is introduced, to obtain the \emph{average} Green's function. This enables us to define the scattering mean free path $\lscat$. With the help of the average Green's function we write down a Bethe-Salpeter equation for the average intensity propagator which is solved within the ladder approximation and describes incoherent (or classical) transport. We close the section with the derivation of a steady state diffusion equation for the average intensity on large length scales. The transport mean free path $\ltrans$ is defined with the help of this equation.
	\item In \emph{\sref{swl}} interference corrections of the mean intensity are considered. We first specify the types of diagrams we include into our calculations, and show how these affect the transport mean free path $\ltrans$. We show that these additional diagrams also require a change of the scattering mean free path $\lscat$. The evaluation of the corrections of $\lscat$ and $\ltrans$ is sketched, and analytical results for the leading order terms are derived.
	\item The results obtained in \sref{swl} are summarised and complemented in \emph{\sref{s:results}}. The leading order contributions to the corrections of the scattering and transport mean free path are discussed, and the results are related to previous studies on disordered electronic systems concerning the weak localisation corrections of the conductivity.
	\item The \emph{last section} concludes this article, and gives a short outlook on future perspectives.
\end{itemize}

\section{Green's functions, disorder average and classical contribution}

After introducing the random wave equation as the starting point of our considerations, and giving the reader a short reminder about diagrammatic techniques for performing the disorder average, we will focus in this chapter on the regime of incoherent transport. Within the ladder approximation, valid for weak disorder, the average intensity is found to follow a classical random walk, with the average step length given by the \emph{scattering mean free path} $\lscat$. On large length scales, the random walk reduces to a diffusion equation, which, as described in the introduction, defines the \emph{transport mean free path}.

\subsection{Model}
We investigate the propagation of classical waves in a three dimensional disordered medium described by the stationary Helmholtz equation
\begin{equation}
	\Bigl(\Delta + k^2\bigl( 1 + V({\bi r}) \bigr) \Bigr) \psi({\bi r}) = j({\bi r}),
	\label{eq:helmholtz}
\end{equation}
where $\psi$ denotes the field, $k$ the wave number, and $j$ is the source term. The potential $V$, describing the disordered medium, is a random function which is assumed to be uncorrelated (white noise) and to exhibit Gaussian statistics:
\begin{equation}
\eqalign{
	\phantom{V({\bi r}')}\langle V({\bi r}) \rangle &\equiv 0 \\
	k^4\langle V({\bi r}) V({\bi r}') \rangle &= \frac{4\pi}{\ell}\delta({\bi r}-{\bi r}').
}
	\label{eq:randompotential}
\end{equation}
Here, $4\pi/\ell$ is a measure for the strength of the potential parametrised by a length scale $\ell$. In the limit of a very weakly disordered medium, the scattering mean free path turns out to be given by $\ell$, see \sref{sec:dilute} below. We use the Green's function method to treat the problem. For the Helmholtz equation with $V\equiv 0$, i.e.\ the vacuum case, the retarded momentum space Green's function $\widetilde{G}_{0}$ is well known to be of the form (see, e.g., \cite{hittmair}, chapter XII.9)
\begin{equation}
	\widetilde{G}_{0}(p) = \frac{1}{k^2 - p^2 + \rmi \eta} \;,
	\label{eq:g0tilde}
\end{equation}
with $\eta$ an infinitesimal positive quantity. In real space, the retarded Green's function then takes the form of a spherical wave
\begin{equation}
	G_{0}(r) = -\frac{\exp(\rmi k r)}{4\pi r} \;,
	\label{eq:g0}
\end{equation}
and the advanced Green's function is the complex conjugate $G_{0}^{*}$.

In the presence of a scattering potential $V$, the Green's function $g$ can be expressed in terms of the vacuum Green's function $G_{0}$ and the potential $V$ by means of the well known \emph{Born series}
\begin{equation}
\eqalign{
	g({\bi r},{\bi r}') &= G_{0}({\bi r},{\bi r}') + \int\rmd{\bi r}_{1} G_{0}({\bi r},{\bi r}_{1})V({\bi r}_{1})G_{0}({\bi r}_{1},{\bi r}')\\
 &+ \int\rmd{\bi r}_{1}\rmd{\bi r}_{2} G_{0}({\bi r},{\bi r}_{1})V({\bi r}_{1})G_{0}({\bi r}_{1},{\bi r}_{2})V({\bi r}_{2})G_{0}({\bi r}_{2},{\bi r}') + \cdots,
}
	\label{eq:born}
\end{equation}
which can be recast into a \emph{Lippmann-Schwinger equation}:
\begin{equation}
	g({\bi r},{\bi r}') = G_{0}({\bi r},{\bi r}') + \int\rmd{\bi r}_{1} G_{0}({\bi r},{\bi r}_{1}) V({\bi r}_{1}) g({\bi r}_{1},{\bi r}').
	\label{eq:lippmannschwinger}
\end{equation}
To shorten the notation one expresses integral equations of this type with the help of diagrams: different types of lines represent the different types of Green's functions ($G_{0}(r)$, $g({\bi r},{\bi r}')$ or, later, the average Green's function $G(r)$), crosses denote the scattering potential. In the next section we will perform the averaging over the disorder potential $V$. The appearing correlators $\langle V({\bi r})V({\bi r}') \rangle$ are represented diagrammatically by crosses connected via a dotted line. As opposed to the Green's function $g$ for a specific realisation of the disorder potential, the average Green's function $G$ is translationally invariant.

\subsection{Disorder average}

The average Green's function $G(\left| {\bi r} - {\bi r}' \right|)$ in presence of the random medium can be expressed via a \emph{Dyson equation} (see, e.g., \cite{hittmair}, chapter XIII.23):
\begin{equation}
	G(\left| {\bi r} - {\bi r}' \right|) = G_{0}(\left| {\bi r} - {\bi r}' \right|) + \int\rmd{\bi r}_{1}\rmd{\bi r}_{2} G_{0}(\left| {\bi r} - {\bi r}_{1}\right|) \Sigma(\left|{\bi r}_{1} - {\bi r}_{2}\right|) G(\left|{\bi r}_{2} - {\bi r}'\right|),
	\label{eq:dyson}
\end{equation}
where $\Sigma(r)$ is the \emph{self energy} defined as the sum of all irreducible diagrams, i.e.\ diagrams which do not fall apart upon cutting a single Green's function, see \fref{fig:sigmau}. From \eref{eq:dyson}, we obtain an algebraic solution for the Fourier transform of the average Green's function $\widetilde{G}$, in terms of the vacuum Green's function $\widetilde{G}_{0}$ and of the self energy $\widetilde{\Sigma}$:
\begin{equation}
	\widetilde{G}(p) = \frac{1}{\widetilde{G}^{-1}_{0}(p) - \widetilde{\Sigma}(p)} = \frac{1}{k^2 - p^2 - \widetilde{\Sigma}(p)} \;.
	\label{eq:gtilde}
\end{equation}
Here, we do not need the infinitesimal displacement of the poles $\eta$ like in \eref{eq:g0tilde}, since the self-energy already possesses a finite imaginary part (see below) which shifts the poles away from the real axis. For weak disorder, i.e.\ $k\ell\gg 1$, the self energy depends only weakly on the momentum $p$ and we can therefore write:
\begin{equation}
	\widetilde{G}(p) = \frac{1}{\tilde{k}^2 - p^2} \;,
	\label{eq:averagegreensfunctiontilde}
\end{equation}
with the effective wave-vector $\tilde{k}=\sqrt{k^2-\Sigma(k)}$. Neglecting the real part of the self energy (which can be absorbed by a suitable redefinition of the wave number $k$), this can be written as $\tilde{k} = k + \rmi/2\lscat$, with the scattering mean free path
\begin{equation}
	\lscat=-\frac{k}{{\rm Im}\tilde{\Sigma}(k)}.
	\label{eq:deflscat}
\end{equation}
In real space, the retarded Green's function takes the form
\begin{equation}
	G(\left| {\bi r}-{\bi r}' \right|) = -\frac{\exp(\rmi \tilde{k} \left| {\bi r} - {\bi r}' \right|)}{4\pi \left| {\bi r} - {\bi r}' \right|}\;,
	\label{eq:averagegreensfunction}
\end{equation}
and the advanced Green's function is again the complex conjugate of \eref{eq:averagegreensfunction}. We see that, due to the disorder, the real space average Green's function decays exponentially on a length scale given by twice the scattering mean free path $\lscat$. Therefore, we will later determine the scattering mean free path as the decay length of the squared absolute value of the average Green's function, see \eref{eq:p0}. In the next section we will see that $\lscat$ is the length scale on which the coherent intensity decays.

To investigate the mean intensity $\mathcal{I}=\langle\psi \psi^*\rangle$ (the brackets denote the average over disorder realisations), one defines the average intensity propagator $\Phi$ as the disorder averaged product of a retarded with an advanced Green's function. The average intensity propagator fulfils the Bethe-Salpeter equation \cite{akkermans:mesoscopic}
\begin{equation}
	\eqalign{
		\fl\Phi({\bi r}_{1}, {\bi r}_{2}; {\bi r}_{3}, {\bi r}_{4}) = G(\left|{\bi r}_{1}-{\bi r}_{3}\right|) G^*(\left|{\bi r}_{2}-{\bi r}_{4}\right|) \\
		+ \int\rmd{\bi r}_{5}\cdots\rmd{\bi r}_{8} \Phi({\bi r}_{1},{\bi r}_{2}; {\bi r}_{5},{\bi r}_{6}) U({\bi r}_{5},{\bi r}_{6}; {\bi r}_{7},{\bi r}_{8}) G(\left|{\bi r}_{7}-{\bi r}_{3}\right|) G^*(\left|{\bi r}_{8}-{\bi r}_{4}\right|) ,
	}
	\label{eq:phi}
\end{equation}
where $U$ denotes the \emph{irreducible vertex} or \emph{intensity operator}, see \fref{fig:sigmau}.

\subsection{Weak disorder approximation}
\label{sec:dilute}

\begin{figure}
	\begin{indented}
	\item[]\includegraphics{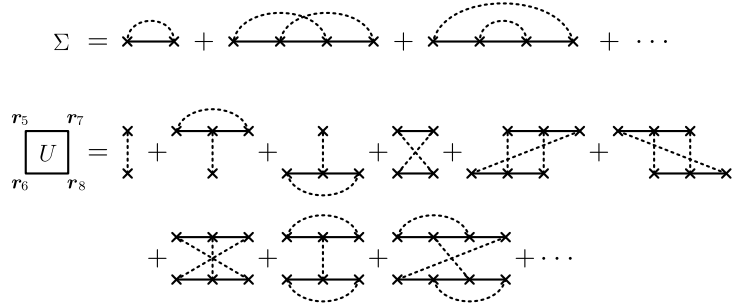}
	\end{indented}
	\caption{Diagrammatic representation of the first terms of the self energy $\Sigma$ and of the irreducible vertex $U$. Crosses connected by a dotted line represent the two-point correlation function $\langle V({\bi r}) V({\bi r}') \rangle$ of the random potential, see \eref{eq:randompotential}. Solid lines correspond to vacuum Green's functions (in $\Sigma$) or average Green's functions (in $U$).}
	\label{fig:sigmau}
\end{figure}
In the case of weak disorder, i.e., $k\ell\gg 1$, one may apply the so-called ladder approximation. Here, only the first term of the self-energy and the irreducible vertex are considered (see \fref{fig:sigmau}). In this approximation, the self-energy and the irreducible vertex read
\begin{eqnarray}
	\Sigma(\left| {\bi r} - {\bi r}' \right|) \approx \frac{4\pi}{\ell} \delta({\bi r} - {\bi r}') G_{0}(\left|{\bi r} - {\bi r}'\right|)
\label{eq:sigmaladder}\\
	U({\bi r}_{5},{\bi r}_{6}; {\bi r}_{7},{\bi r}_{8}) \approx \frac{4\pi}{\ell}\delta({\bi r}_{5}-{\bi r}_{6})\delta({\bi r}_{5}-{\bi r}_{7})\delta({\bi r}_{6}-{\bi r}_{8}).\label{eq:uladder}
\end{eqnarray}
Together with \eref{eq:deflscat}, this approximation yields $\lscat=\ell$. Using $U$ of the form \eref{eq:uladder} in the Bethe-Salpeter equation \eref{eq:phi}, and assuming a $\delta$-source of strength $s_0$ at ${\bi r}_{0}$, i.e.\ $j({\bi r})=\sqrt{s_{0}}\delta({\bi r}-{\bi r}_{0})$ in \eref{eq:helmholtz}, then leads to the following transport equation for the mean intensity $\mathcal{I}({\bi r}) = \Phi({\bi r}_{0},{\bi r}_{0};{\bi r},{\bi r})$:
\begin{equation}
	\mathcal{I}({\bi r}) = \mathcal{I}_{0}({\bi r}) + \int\rmd{\bi r}' P_{0}(\left|{\bi r}-{\bi r}'\right|)\, \mathcal{I}({\bi r}'),
	\label{eq:averageintensity}
\end{equation}
where $\mathcal{I}_{0}({\bi r})=s_{0}\left| G(\left|{\bi r}_{0}-{\bi r}\right|) \right|^{2}$ denotes the coherent intensity, i.e., the intensity arriving at point ${\bi r}$ without being scattered, and
\begin{equation}
	P_{0}(\left| {\bi r}-{\bi r}' \right|) = \frac{4\pi}{\ell}\left| G(\left| {\bi r}-{\bi r}' \right|) \right|^2 = \frac{\exp(-\left| {\bi r}-{\bi r}' \right|/\lscat)}{4\pi\ell \left| {\bi r}-{\bi r}' \right|^2}\,.
	\label{eq:p0}
\end{equation}
Although $\lscat$ equals $\ell$ within the ladder approximation, we still keep $\lscat$ in the exponential since, later, when additional interference diagrams are taken into account, this equality will not hold any longer and a distinction between $\lscat$ and $\ell$ is crucial. As mentioned in the introductory \sref{s:introduction}, the coherent intensity decays exponentially on the scale of the scattering mean free path, which can be seen from \eref{eq:averagegreensfunction} together with \eref{eq:deflscat}.

The transport equation \eref{eq:averageintensity} can be interpreted in terms of a classical random walk. According to \eref{eq:averageintensity}, this random walk is characterized by the source term $\mathcal{I}_{0}({\bi r})$ and the probability distribution $P_{0}(\left| {\bi r}-{\bi r}' \right|)$ of the step length $\left| {\bi r}-{\bi r}' \right|$ between two successive scattering events at $\bi r$ and $\bi r'$. Remember that this simple form of the transport equation only applies to isotropic scattering from a $\delta$-correlated potential. In the case of anisotropic scattering (i.e., if the random potential exhibits correlations on a length scale comparable to the wavelength), one can derive, instead of \eref{eq:averageintensity}, a radiative transfer equation for the intensity $\mathcal{I}({\bi r},\hat{{\bi s}})$ at point ${\bi r}$, with propagation direction $\hat{{\bi s}}$ \cite{ishimaru:waveprop}.

\begin{figure}
	\begin{indented}
	\item[]\includegraphics{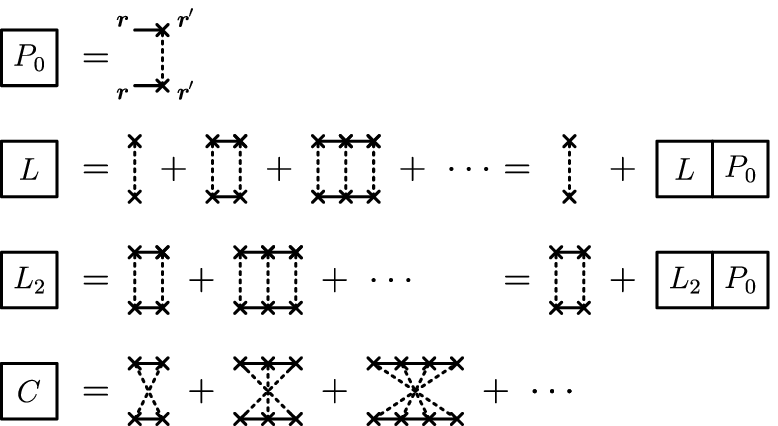}
	\end{indented}
	\caption{Diagrammatic representation of the ``single step'' $P_{0}$, the ladder and crossed propagators $L$, $L_{2}$ and $C$, and of the Bethe-Salpeter equation \eref{eq:bethesalpeterladder} satisfied by the ladder propagator. By reciprocity symmetry, $C$ and $L_{2}$ have the same value, see \eref{eq:cnlm}.
	}
	\label{fig:props}
\end{figure}

The solution of the transport equation \eref{eq:averageintensity} can be expressed in terms of the \emph{ladder propagator} $L(\left|{\bi r}_1-{\bi r}_2\right|)$, which we will need as a building block for the weak localisation diagrams in \sref{swl}:
\begin{equation}
\mathcal{I}({\bi r}) = \int\rmd{\bi r}_1\rmd{\bi r}_2 \left| G(\left|{\bi r}-{\bi r}_2\right|) \right|^{2} L(\left|{\bi r}_1-{\bi r}_2\right|)\, \mathcal{I}_0({\bi r}_1)\, .
\end{equation}
Equivalently to \eref{eq:averageintensity}, the transport equation for the ladder propagator reads:
\begin{equation}
	L(|{\bi r}_1-{\bi r}_2|) = \frac{4\pi}{\ell}\delta({\bi r}_1-{\bi r}_2)+ \int\rmd{\bi r}_3 L(|{\bi r}_1-{\bi r}_3|) P_{0}(\left|{\bi r}_3-{\bi r}_2\right|)\, ,
	\label{eq:bethesalpeterladder}
\end{equation}
see \fref{fig:props}. In Fourier space, the sequence of single steps reduces to a geometric series:
 \begin{equation}
	\widetilde{L}(q) = \frac{4\pi}{\ell} \sum_{n=0}^\infty \left(\widetilde{P}_0(q)\right)^n=
	\frac{4\pi}{\ell}\frac{1}{1 - \frac{\lscat}{\ell}\frac{\arctan(q\lscat)}{q\lscat}} \;,
	\label{eq:ladderprop}
\end{equation}
where we used the Fourier transform $\widetilde{P}_{0}$ of \eref{eq:p0}
\begin{equation}
	\widetilde{P}_{0}(q) = \frac{\lscat}{\ell} \frac{\arctan(q\lscat)}{q\lscat} \;.
	\label{eq:p0tilde}
\end{equation}
Summing up all ladder diagrams without the first $({\rm N}-1)$ terms leads to the ladder propagator with at least ${\rm N}$ scattering events:
\begin{equation}
	\widetilde{L}_{\rm N}(q) = \frac{4\pi}{\ell} \sum_{n={\rm N-1}}^\infty \left(\widetilde{P}_0(q)\right)^n= \frac{4\pi}{\ell}\frac{\left( \frac{\lscat}{\ell}\arctan(q\lscat)/q\lscat \right)^{{\rm N}-1}}{1 - \frac{\lscat}{\ell}\arctan(q\lscat)/q\lscat}.
	\label{eq:lnprop}
\end{equation}
Thus, $L_{1} \equiv L$ holds.

Similarly, we define the crossed propagator $C$ as the sum of all \emph{maximally crossed} diagrams \cite{langerneal:prl16} (see \fref{fig:props}), for later use. By reciprocity symmetry \cite{tiggelen:1997}, $C$ equals $L_{2}$. Furthermore, we can construct propagators $C_{\rm N}$ consisting of maximally crossed diagrams without the first $({\rm N}-1)$ terms. It then holds that $C_{1}\equiv C$, and
\begin{equation}
	\widetilde{C}_{\rm N}(q) = \widetilde{L}_{\rm N+1}(q) \qquad {\rm N}=1,2,3,\dots \;,
	\label{eq:cnlm}
\end{equation}
generally. Diagrammatic representations of the propagators $P_{0}$, $L$, $L_{2}$ and $C$ are given in \fref{fig:props}.

\subsection{Diffusion approximation \& transport mean free path}\label{sec:diffusion}
If we consider transport on large length scales (compared to $\lscat$), we may derive a \emph{steady state diffusion equation} starting from \eref{eq:averageintensity}. For this, we perform a Taylor expansion of the mean intensity $\mathcal{I}({\bi r}')$ around position ${\bi r}$ inside the integral in \eref{eq:averageintensity}:
\begin{equation}
	\mathcal{I}({\bi r}+{\bi r}') = \mathcal{I}({\bi r}) + {\bi r}'\cdot\nabla\mathcal{I}({\bi r}) + \frac{1}{2} r_i' r_j'\frac{\partial^2\mathcal{I}({\bi r})}{\partial r_i \partial r_j} + \cdots ,
	\label{eq:taylor}
\end{equation}
where summation over repeated indices is implied. Stopping the expansion at the second order and insertion into \eref{eq:averageintensity} yields
\begin{equation}
	\mathcal{I}({\bi r}) = \mathcal{I}_{0}({\bi r}) + \int\rmd{\bi r}' P_{0}({\bi r}') \left( \mathcal{I}({\bi r}) + {\bi r}'\cdot\nabla\mathcal{I}({\bi r}) + \frac{1}{2} r_i' r_j'\frac{\partial^2\mathcal{I}({\bi r})}{\partial r_i \partial r_j} \right).
	\label{eq:averageintensitytaylor}
\end{equation}
The first term in the integral cancels the left hand side of the equation, because of the normalisation of $P_{0}$:
\begin{equation}
	\int\rmd{\bi r} P_{0}(r) = 1\;.
	\label{eq:normalisation}
\end{equation}
The second term vanishes because it is an odd function of the integration coordinates. Thus, only the term quadratic in ${\bi r}'$ survives inside the integral:
\begin{equation}
	0 = \mathcal{I}_{0}({\bi r}) + \frac{1}{2}\int\rmd{\bi r}' P_{0}({\bi r}')\; r_i' r_j'\frac{\partial^2\mathcal{I}({\bi r})}{\partial r_i \partial r_j} \;.
	\label{eq:survivor}
\end{equation}
By exploiting the spherical symmetry of $P_{0}$, we can use the identity
\begin{equation}
	\int\rmd{\bi r}' P_{0}({\bi r}') {r_{i}'}^{2} = \int\rmd{\bi r}' P_{0}({\bi r}') {r'_{j}}^{2} \qquad ,\quad\forall i,j
	\label{eq:p0r2}
\end{equation}
and conclude that
\begin{equation}
	\int\rmd{\bi r}' P_{0}({\bi r}')\; r_i' r_j'\frac{\partial^2\mathcal{I}({\bi r})}{\partial r_i \partial r_j} = \frac{1}{3} \int\rmd{\bi r}' P_{0}({\bi r}') {\bi r}'^{2} \Delta \mathcal{I}({\bi r}).
	\label{eq:laplace}
\end{equation}
where $\Delta=\nabla^2$ denotes the Laplace operator. Thus \eref{eq:survivor} takes the form of a stationary diffusion equation:
\begin{equation}
	\mathcal{D}^{\rm (s)}\Delta\mathcal{I}({\bi r}) + \mathcal{I}_{0}({\bi r}) = 0.
	\label{eq:fundamentalsteadystatediffusionequationreduced}
\end{equation}
It only depends on a single parameter $\mathcal{D}^{\rm (s)}$ which is the stationary analogue of the diffusion coefficient appearing in the time dependent diffusion equation. The parameter $\mathcal{D}^{\rm (s)}$ is given by the relation
\begin{equation}
	\mathcal{D}^{\rm (s)} = \frac{1}{6}\int\rmd{\bi r}\,r^{2}P_{0}(r).
	\label{eq:variance}
\end{equation}
By comparison with the fundamental steady state diffusion equation, found in \cite{ishimaru:waveprop} chapter 9, we see that
\begin{equation}
	\mathcal{D}^{\rm (s)} = \frac{\lscat\ltrans}{3}\;,
	\label{eq:steadyd}
\end{equation}
which we use to calculate the transport mean free path. The interpretation of $P_{0}$ as the step length distribution (cf.\ \sref{sec:dilute} above) then means that we calculate the transport mean free path via the \emph{variance} of the step length distribution or, equivalently, by the \emph{mean square displacement}.

Within the ladder approximation, we find from \eref{eq:normalisation}, together with \eref{eq:p0},
\begin{equation}
	1 = \frac{4\pi}{\ell}\int\rmd{\bi r} \left| G(r) \right|^2 = \int\rmd{\bi r} \frac{\exp\left( -r/\lscat \right)}{4\pi\ell r^2} = \frac{\lscat}{\ell}\;,
	\label{eq:lscateql}
\end{equation}
and thus $\lscat\equiv\ell$, which is consistent with the result found from \eref{eq:sigmaladder} using the definition \eref{eq:deflscat}. Furthermore, inserting \eref{eq:p0} into \eref{eq:variance} yields $\mathcal{D}^{\rm (s)}=\lscat^2/3 \equiv \ell^2/3$. Combining this with \eref{eq:steadyd} we obtain:
\begin{equation}
	\ltrans \equiv \ell \equiv \lscat\;.
	\label{eq:lll}
\end{equation}
Again, remember that this is only true for isotropic scattering -- otherwise the scattering and transport mean free path would differ already in the ladder approximation.

To close this section, it is instructive to look at the solution of \eref{eq:fundamentalsteadystatediffusionequationreduced} for a $\delta$-source of strength $s_0$, i.e., $\mathcal{I}_0({\bi r})=s_0\left| G(\left|{\bi r}\right|) \right|^{2}=s_0\exp(-r/\lscat)/(4\pi r)^2$ in an infinite medium. For $r\gg\lscat$, the exponentially localized coherent intensity can be replaced by $\mathcal{I}_{0}({\bi r}) = s_{0}\lscat\delta\left( {\bi r} \right)/4\pi$, for which the solution of \eref{eq:fundamentalsteadystatediffusionequationreduced} reads
\begin{equation}
	\mathcal{I}(r) = \frac{3 s_{0}}{16\pi^2\ltrans r}\;.
\label{eq:intensitydeltasource}
\end{equation}
This is the asymptotic $1/r$ decay mentioned in \sref{s:introduction}, with the proportionality constant $3s_{0}/16\pi^2\ltrans$ thus defining the transport mean free path. For given source strength $s_0$, a measurement of the transport mean free path can therefore be performed via the detection of the average intensity at a large distance $r$ from the source (where the coherent contribution is sufficiently attenuated).

\section{Weak disorder corrections}
\label{swl}

In the previous section, we saw that, for isotropic scattering and within the ladder approximation, scattering and transport mean free path are equal, \eref{eq:lll}. However, in the presence of reciprocity symmetry, additional diagrams, apart from the ladders, play a role. These contributions will give rise to corrections of the mean free paths and will therefore lead to deviations from \eref{eq:lll}.

In this section, we will introduce these additional diagrams and show how they are incorporated into the transport equation for the average intensity. We will then calculate the corrections due to these diagrams, first for the scattering and then for the transport mean free path, and identify the leading contributions in the disorder parameter $1/(k\ell)$.

\subsection{Contributing diagrams}

\begin{figure}
	\begin{indented}
	\item[]\includegraphics{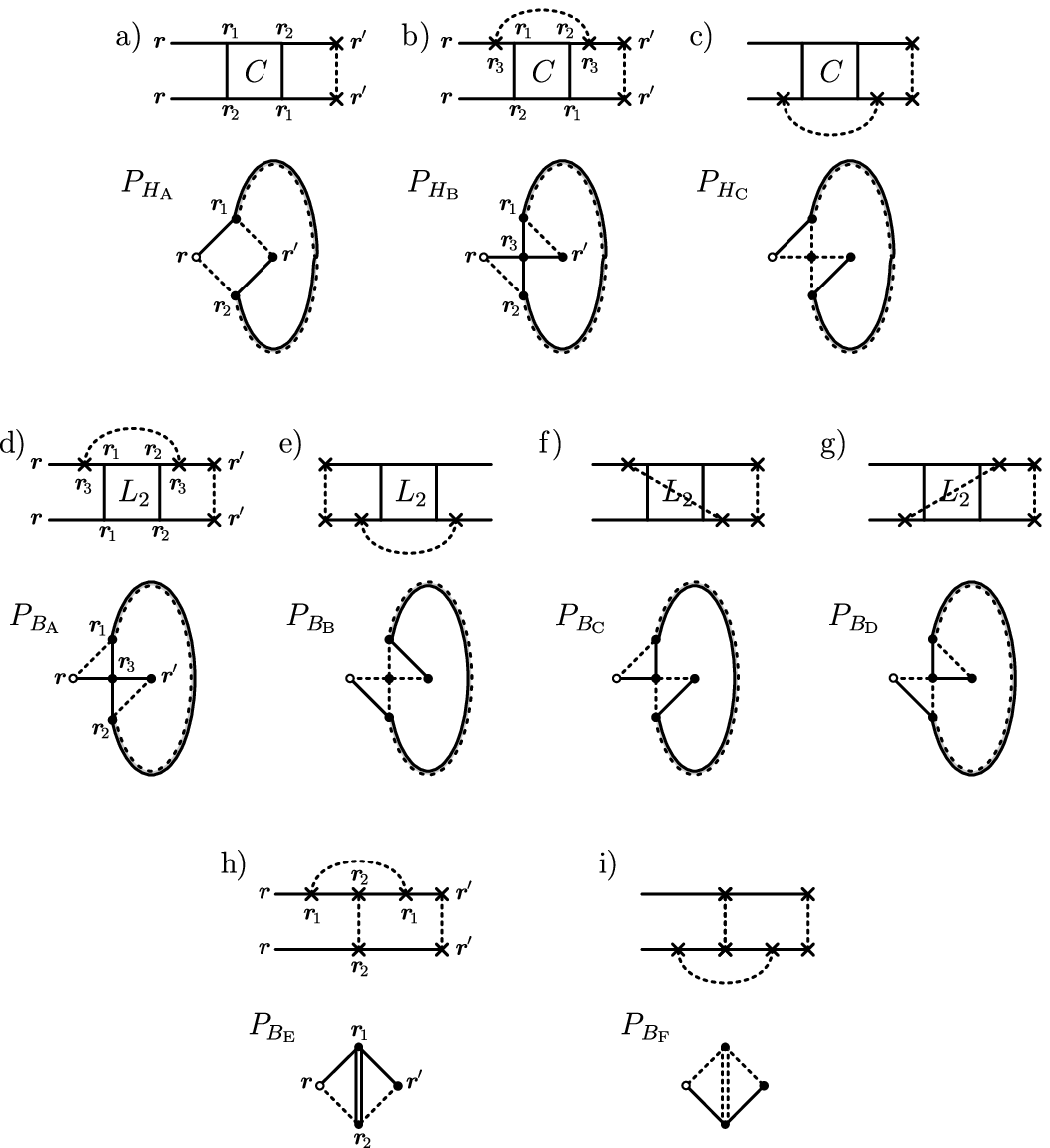}
	\end{indented}
	\caption{Diagrammatic representations of the weak localisation contributions to the step length distribution $P(\left| {\bi r}-{\bi r}' \right|)$. Each diagram is depicted in two different but equivalent representations: In the upper diagrams, the upper (lower) horizontal lines denote retarded (advanced) Green's functions, crosses connected by dashed lines refer to the two-point correlation function of the potential, see \eref{eq:randompotential} -- which, in turn, amounts to a scattering event -- and $L_{2}$ and $C$ represent the ladder and crossed propagators defined in \eref{fig:props}. In the lower type diagrams, solid (or dashed) lines denote average Green's functions $G$ (or $G^{*}$), full circles refer to scattering events, open circles to a specific point in space (without scattering event), whereas paired solid/dashed lines, (\protect\includegraphics{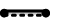}) or (\protect\includegraphics{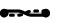}), indicate ladder or crossed propagators, respectively.}
	\label{fig:diagrams}
\end{figure}

We start by specifying the types of diagrams we include into the irreducible vertex $U$. Obviously, we have to consider the ladder contribution, see \eref{eq:uladder}, which accounts for classical, i.e.\ interference-free, propagation. Interference-free means that the two amplitudes which constitute the intensity are perfectly in phase at any instance of the scattering sequence. Since, in the presence of reciprocity symmetry, the two amplitudes forming a maximally crossed diagram do not collect any phase difference either, these have to be included as well, see \eref{eq:cnlm}. When a maximally crossed diagram is connected with ladder diagrams on both sides, a crossing is generated. This crossing introduces a phase difference between the involved amplitudes and thus vanishes in the limit of $k\ell\rightarrow\infty$, i.e., the corresponding diagram will yield a small contribution in the weak disorder limit. A crossing can be decorated with one two-point potential correlation function $\langle V({\bi r})V({\bi r}') \rangle$ \eref{eq:randompotential}, i.e.\ crosses connected with a dotted line, without rendering the diagram higher order in the disorder parameter $1/k\ell$. In the case of the maximally crossed diagrams, these crossings are known in the literature as four point \emph{Hikami-boxes} \cite{hikami:prb24}. Similar structures appear if one additional correlation function is introduced inside ladder diagrams, cf.\ \fref{fig:diagrams} d) - i), whereas adding more than one of these correlation functions leads to contributions of higher order \cite{akkermans:mesoscopic,wysokinski:pre52}. Since any of these irreducible vertices, depicted in \fref{fig:diagrams}, may be inserted between two ladder scattering events at $\bi r$ and $\bi r'$, they will lead to a modified step length distribution $P(|{\bi r}-{\bi r'}|)$ (see below). The diagrammatic representation in \fref{fig:diagrams} is given in two different, but equivalent ways: the upper one is already familiar from \fref{fig:sigmau} and \fref{fig:props}, whereas the lower one corresponds to a real-space representation emphasizing especially the loop-like character of the additional diagrams. The diagrams carrying an index $H$ are the familiar Hikami-boxes \cite{hikami:prb24}. Since our aim in this article is to calculate the \emph{leading} order weak localisation corrections, we do not consider a whole group of ``small'' diagrams (e.g.\ the diagrams given by the last two terms, the eighth and the ninth, of the irreducible vertex depicted in \fref{fig:sigmau}) which contribute in order $1/(k\ell)^2$ and which can be found in \cite{wysokinski:pre52}. With this selection of the contributing diagrams we completely account for the corrections in $1/k\ell$ and $\ln(k\ell)/(k\ell)^2$ completely, see \sref{s:results}. The mean intensity can be constructed from all possible combinations of these building blocks, together with the ladder building block $P_0$ (figure \ref{fig:props}). This is achieved with a Bethe-Salpeter equation like \eref{eq:averageintensity}
\begin{equation}
	\mathcal{I}({\bi r}) = \mathcal{I}_{0}({\bi r}) + \int\rmd{\bi r}' P(\left|{\bi r}-{\bi r}'\right|)\, \mathcal{I}({\bi r}'),
	\label{eq:averageintensitymod}
\end{equation}
where, now, in comparison to \eref{eq:averageintensity}, $P_0$ is replaced by the modified step length distribution $P$ which, apart from $P_{0}$, contains the additional diagrams shown in \fref{fig:diagrams}, see \eref{eq:steplength} below.

Our approach is very similar to the one used in \cite{assaf:prb63}, where the weak localisation corrections of spectral statistical properties, namely the two level correlation function, of two-dimensional disordered conductors were calculated for systems with spatial extent of the order of the scattering mean free path (denoted as ``ballistic limit'', there). These authors also use the exact propagator within the loops and the exact form of the Hikami boxes. In \cite{assaf:prb63}, the model considered is a non-interacting electron gas on a \emph{two dimensional torus}. The use of a finite medium is necessary, since in a two dimensional infinite medium all states are localised. Thus, the authors study the behaviour of the weak localisation corrections with respect to changes of the medium size. In contrast, we perform our calculations for a \emph{three dimensional infinite} medium below the mobility edge, i.e.\ in the weak localisation regime. We analyse the weak localisation corrections of transport quantities, like the transport mean free path, under changes of the \emph{disorder strength}. In the view of \cite{assaf:prb63}, our approach is to allow the loops, which lead to the weak localisation corrections by influencing the return probability, to also probe short scattering paths -- meaning the loops not only explore the ``diffusive'', but also the ``ballistic'' scale of the medium.

Lets return to the modified step length distribution $P$. Summing up all diagrams shown in \fref{fig:diagrams} yields
\begin{equation}
	P({\bi r}) = P_0({\bi r}) + P_{H}({\bi r}) + P_{B}({\bi r}),
	\label{eq:steplength}
\end{equation}
where
\begin{equation}
	P_{H} = P_{H_{\rm A}} + P_{H_{\rm B}} + P_{H_{\rm C}} \\
	\label{eq:defsph}
\end{equation}
and
\begin{equation}
	P_{B} = P_{B_{\rm A}} + P_{B_{\rm B}} + P_{B_{\rm C}} + P_{B_{\rm D}} + P_{B_{\rm E}} + P_{B_{\rm F}}.
	\label{eq:defspl}
\end{equation}
The respective diagrams fulfil the following symmetry relations: mirroring vertically is equivalent to complex conjugation (interchange of the upper line, $G$, and the lower line, $G^*$), whereas mirroring horizontally (either the complete diagram or just the upper or lower half) does not change the value of the diagram (reciprocity symmetry). Thereby, we can infer
\begin{equation}
	P_{H_{\rm B}}=P_{H_{\rm C}}^*=P_{B_{\rm A}}=P_{B_{\rm B}}^*, \quad
	P_{B_{\rm C}}=P_{B_{\rm D}}^*, \quad
	P_{B_{\rm E}}=P_{B_{\rm F}}^*,\label{eq:symmetry}
\end{equation}
i.e.\ only four different diagrams have to be calculated: $P_{H_{\rm A}}$, $P_{H_{\rm B}}$, $P_{B_{\rm C}}$ and $P_{B_{\rm E}}$.

As already pointed out before, the inclusion of additional diagrams into the step length distribution $P$ entails additional contributions to the irreducible vertex $U$ as compared to the ladder approximation. Since the irreducible vertex and the self-energy are linked via a Ward identity which ensures conservation of the total energy flux \cite{vollhardt:prb22}, this also implies the inclusion of additional diagrams, apart from \eref{eq:sigmaladder}, into the self-energy. This results in a change of the scattering mean free path. In the case of \eref{eq:averageintensitymod}, energy flux conservation is ensured by the normalisation of the modified step length distribution
\begin{equation}
	\int\rmd{\bi r} P(r) = 1,
	\label{eq:normalisationfullp}
\end{equation}
from which we will calculate $\lscat$ as shown below. As we have checked, the same result is obtained when calculating $\lscat$ directly from the self-energy, see \eref{eq:deflscat}, using a Ward identity \cite{eckert:diplom}. On the other hand, the variance of the step length distribution yields the transport mean free path via
\begin{equation}
	\frac{1}{6}\int\rmd{\bi r}\; r^2 P(r) = \frac{\lscat\ltrans}{3}\;,
	\label{eq:variancefullp}
\end{equation}
as discussed above, cf.\ \eref{eq:steadyd}.

Finally, we give an example of how to translate the diagrams depicted in \fref{fig:diagrams} into formulas. The diagram \ref{fig:diagrams} b), e.g., represents the equation
\begin{equation}
	\eqalign{
		\fl P_{H_{\rm B}}(\left| {\bi r} - {\bi r}' \right|) = \left( \frac{4\pi}{\ell} \right)^{2} \int\rmd{\bi r}_{1}\rmd{\bi r}_{2}\rmd{\bi r}_{3} G(\left| {\bi r}-{\bi r}_{3} \right|) G(\left| {\bi r}_{3}-{\bi r}_{1} \right|) G^{*}(\left| {\bi r}-{\bi r}_{2} \right|)\\
		C(\left| {\bi r}_{1}-{\bi r}_{2} \right|) G(\left| {\bi r}_{2}-{\bi r}_{3} \right|) G(\left| {\bi r}_{3}-{\bi r}' \right|) G^{*}(\left| {\bi r}_{1}-{\bi r}' \right|).
	}
	\label{eq:phbexample}
\end{equation}
We recognize the four Green's functions $G$, the two complex conjugate Green's functions $G^*$, and the crossed propagator $C$. According to its definition, see \eref{eq:cnlm} and \fref{fig:props}, $C$ already contains two scattering events at the starting and end points (${\bi r}_{1}$ and ${\bi r}_{2}$, respectively). The remaining two scattering events (at ${\bi r}_{3}$ and ${\bi r}'$) are taken into account by the prefactor $(4\pi/\ell)^{2}$ originating from the potential correlation function \eref{eq:randompotential}. We \emph{must not} include a factor of $4\pi/\ell$ for a scattering event at ${\bi r}$, since the transport equation \eref{eq:averageintensitymod} would otherwise double count these events upon concatenating the building blocks.

\subsection{The scattering mean free path}\label{s:lscat}
Since, according to \eref{eq:averagegreensfunction}, $\lscat$ gives the decay length of the average Green's function, it follows from \eref{eq:p0} that
\begin{equation}
	\int\rmd{\bi r} P_{0}(r) = \frac{\lscat}{\ell} \;.
	\label{eq:lsoverl}
\end{equation}
The scattering mean free path is thus obtained from the normalisation \eref{eq:normalisationfullp} of the total step length distribution $P=P_0+P_H+P_B$, \eref{eq:steplength}, as follows:
\begin{equation}
	\frac{\lscat}{\ell} = 1 - \int\rmd{\bi r} \Bigl( P_{H}({\bi r}) + P_{B}({\bi r}) \Bigr) \equiv 1 + \frac{\delta H_{\rm s}}{\ell} + \frac{\delta B_{\rm s}}{\ell}\;,
	\label{eq:loverls}
\end{equation}
where we introduced the corrections of the scattering mean free path due to the different Hikami, $\delta H_{\rm s}$, and dressed ladder boxes, $\delta B_{\rm s}$. To evaluate the integrals, we switch to Fourier space, since there we know the analytic form of the ladder/crossed propagator $\widetilde{L}/\widetilde{C}$, see \eref{eq:lnprop} and \eref{eq:cnlm}. The corrections thus take the form (beware the minus sign from \eref{eq:loverls}!)
\begin{equation}
	\eqalign{
		\frac{\delta H_{\rm s}}{\ell} &= -\int\rmd{\bi r}\rmd{\bi r}'\rmd{\bi r_{12}}\, H({\bi r},{\bi r_1},{\bi r}',{\bi r_2}) C(r_{12}) \\
		&= -\int\rmd{\bi r_{12}}\, H(r_{12}) C(r_{12}) \\
		&= -\int\frac{\rmd{\bi q}}{(2\pi)^3}\widetilde{H}(q) \widetilde{C}(q).
	}
	\label{eq:deltah}
\end{equation}
Here, $H({\bi r},{\bi r}_{1},{\bi r}',{\bi r}_{2})$ represents the sum of diagrams a)--c) (\fref{fig:diagrams}) without the crossed propagators connecting ${\bi r}_{1}$ and ${\bi r}_{2}$. As an example, the contribution from b) reads:
\begin{equation}
	\eqalign{
		\fl H_{\rm B}({\bi r},{\bi r}_{1},{\bi r}',{\bi r}_{2}) = \left( \frac{4\pi}{\ell} \right)^{2} \int\rmd{\bi r}_{3} G(\left| {\bi r}-{\bi r}_{3} \right|) G(\left| {\bi r}_{3}-{\bi r}_{1} \right|) G^{*}(\left| {\bi r}-{\bi r}_{2} \right|)\\
		G(\left| {\bi r}_{2}-{\bi r}_{3} \right|) G(\left| {\bi r}_{3}-{\bi r}' \right|) G^{*}(\left| {\bi r}_{1}-{\bi r}' \right|).
	}
	\label{eq:hbex}
\end{equation}
Integration over the outer points ${\bi r}$ and ${\bi r}'$ yields the reduced Hikami-box $H({\bi r}_{1},{\bi r}_{2})=H(r_{12})$ with Fourier transform $\widetilde{H}(q) = \widetilde{H}_{\rm A}(q) + \widetilde{H}_{\rm B}(q) + \widetilde{H}_{\rm C}(q)$. Inserting this into \eref{eq:deltah} yields, correspondingly, $\delta H_s=\delta H_{\rm A,s}+\delta H_{\rm B,s}+\delta H_{\rm C,s}$. In the same way, we get for the dressed ladder boxes $B_{\rm A,\dots,D}$ (the contributions from $B_{\rm E,F}$ are special and we consider them separately in a moment):
\begin{equation}
	\frac{\delta B_{\rm A+\cdots+D,s}}{\ell} = -\int\frac{\rmd{\bi q}}{(2\pi)^3}\widetilde{B}(q) \widetilde{L}_{2}(q),
	\label{eq:deltal}
\end{equation}
with the Fourier transform of the dressed ladder boxes $\widetilde{B}(q) = \widetilde{B}_{\rm A}(q)+\cdots+\widetilde{B}_{\rm D}(q)$. The Fourier transforms of the boxes are readily obtained to be
\numparts
\begin{equation}
	\widetilde{H}_{\rm A}(q) = \frac{\lscat^2}{4 k^2\ell q}\Bigl( 2\arctan(q\lscat) - \arctan(q\lscat+2k\lscat) - \arctan(q\lscat - 2 k\lscat) \Bigr)
	\label{eq:haq}
\end{equation}
\begin{equation}
	\eqalign{
		\widetilde{H}_{\rm B}(q) =& \left( \frac{1}{4 k q} \right)^2 \Bigl( -2\rmi \arctan(q\ell) \\
			& -\ln(1-\rmi(q\ell-2k\ell)) + \ln(1+\rmi(q\ell+2k\ell)) \Bigr)^2 \\
		\phantom{\widetilde{H}_{\rm B}(q)\!}=& \widetilde{H}^{*}_{\rm C}(q) = \widetilde{B}_{\rm A}(q) = \widetilde{B}^{*}_{\rm B}(q),
	}
	\label{eq:hbqlaq}
\end{equation}
and, finally,
\begin{equation}
	\eqalign{
		\widetilde{B}_{\rm C}(q) =& \left( \frac{1}{4 k q} \right)^2\\
		&\times \bigl[ 2\rmi \arctan(q\ell) + \ln(-\rmi + 2 k\ell - q\ell) - \ln(-\rmi + 2 k\ell + q\ell) \bigr]\\
		&\times \bigl[ -2\rmi \arctan(q\ell) + \ln(\rmi + 2 k\ell - q\ell) - \ln(\rmi + 2k\ell + q\ell) \bigr]\\
		\phantom{\widetilde{B}_{\rm C}(q)\!}=& \widetilde{B}_{\rm D}(q).
	}
	\label{eq:lcqldq}
\end{equation}
\endnumparts
Here, we keep the distinction between $\lscat$ and $\ell$ only in the $H_{\rm A}$-box \eref{eq:haq}, whereas we set $\lscat\equiv\ell$ in the remaining boxes $H_{\rm B,C}$ and $B_{\rm A,B,C,D}$. As we will see later, the latter boxes lead to higher order corrections as compared to $H_{\rm A}$, such that the difference between $\lscat$ and $\ell$ can be neglected. Furthermore, we used the symmetry properties discussed earlier, leading to $\widetilde{H}_{\rm B} = \widetilde{H}^{*}_{\rm C} = \widetilde{B}_{\rm A} = \widetilde{B}^{*}_{\rm B}$, see \eref{eq:symmetry}. The final integrations over $q$, including the propagators $\widetilde{L}_{2}$ and $\widetilde{C}$, see \eref{eq:deltah} and \eref{eq:deltal}, have to be performed numerically. The results will be presented in \sref{s:results}.

The corrections due to the diagrams from \fref{fig:diagrams} h) and i) which do not contain full ladder or crossed propagators can be calculated directly (without going to momentum space), and analytically. Because of translational symmetry, we set ${\bi r}=0$ and obtain ($r_{1/2}=\left| {\bi r}_{1/2} \right|$)
\begin{equation}
	\eqalign{
		\frac{\delta B_{\rm E,s}}{\ell} &= -\left( \frac{4\pi}{\ell} \right)^{3}\int\rmd{\bi r}'\rmd{\bi r}_{1}\rmd{\bi r}_{2}\, G(r_{1}) G(\left| {\bi r}_{1}-{\bi r}_{2} \right|) G^{*}(r_{2})\\
		&\phantom{=\int} G(\left| {\bi r}_{2}-{\bi r}_{1} \right|) G(\left| {\bi r}_{1}-{\bi r}' \right|) G^{*}(\left| {\bi r}_{2}-{\bi r}' \right|)\\
		&= -\frac{\rmi}{2(k\ell)^2}\biggl[ 2k\ell \,{\rm arctanh}\left(\frac{k\ell}{\rmi+k\ell}\right) + (\rmi + k\ell) \ln\left( \frac{-1 + 2\rmi k\ell}{\left( \rmi + k\ell \right)^2} \right) \biggr] \\
		&= \left( \frac{\delta B_{\rm F,s}}{\ell} \right)^{*} \;.
	}
	\label{eq:corrlscatlef}
\end{equation}
For $k\ell\gg 1$, the sum of these two contributions has the asymptotic behaviour
\begin{equation}
	\frac{\delta B_{\rm E+F,s}}{\ell} \equiv \frac{\delta B_{\rm E,s}+\delta B_{\rm F,s}}{\ell} \underset{k\ell\gg 1}{\longrightarrow} \frac{-1 + \ln(2) - \ln(k\ell)}{(k\ell)^2} \;.
	\label{eq:corrlscatlefasymptotic}
\end{equation}

\begin{figure}
	\centering\includegraphics{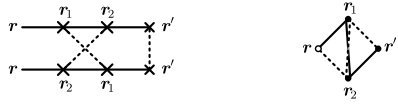}
	\caption{Short-loop diagram responsible for the leading correction of the scattering and transport mean free paths, depicted in two different but equivalent representations (as in \fref{fig:diagrams}). It is obtained as a part of diagram \ref{fig:diagrams} a), where the loop between ${\bi r}_{1}$ and ${\bi r}_{2}$ is replaced by the direct connection, or, in other words, only the first term in the series for the crossed propagator $C$, see \fref{fig:props}, is taken into account.}
	\label{fig:leadingdiagram}
\end{figure}

Moreover, we can obtain the leading correction of the scattering mean free path analytically. It turns out to be given by a part of diagram \ref{fig:diagrams} a), depicted in \fref{fig:leadingdiagram}, where only the first term ($4\pi P_{0}/\ell$) in the series forming the crossed propagator is taken into account:
\begin{equation}
	\eqalign{
		\frac{\delta H_{\rm A,1,s}}{\ell} &= - \int\frac{\rmd{\bi q}}{\left( 2\pi \right)^{3}} \widetilde{H}_{\rm A}(q) \frac{4\pi}{\ell}\widetilde{P}_{0}(q)\\
		&= -\frac{\lscat^2\arctan(k\lscat)}{k\ell^3} + \frac{\lscat\ln(1+(k\lscat)^2)}{2k^2\ell^3} \underset{k\ell\gg 1}{\longrightarrow} -\frac{\pi}{2k\ell} + \frac{\ln\left( k\ell \right)}{\left( k\ell \right)^2}\;.
	}
	\label{eq:lscatha1}
\end{equation}
Again, we set $\lscat\equiv\ell$ to obtain the two leading contributions on the right hand side of \eref{eq:lscatha1}, since $\lscat\rightarrow\ell$ for $k\ell\rightarrow \infty$. (The distinction between $\lscat$ and $\ell$ will affect only the next leading contribution proportional to $1/(k\ell)^2$, see the coefficient $a_{2}$ of $\delta H_{\rm A,1}$ in \tref{tab:coefflscat}, below.)

The remaining contributions to diagram \ref{fig:diagrams} a) are given by replacing $4\pi P_{0}/\ell$ with $C_{2}=L_{3}$, see \eref{eq:cnlm}, in \eref{eq:deltah}, i.e.\ by subtracting a single scattering event from the full $C$:
\begin{equation}
	\frac{\delta H_{\rm A,2,s}}{\ell} = -\int\frac{\rmd{\bi q}}{(2\pi)^3}\widetilde{H}_{\rm A}(q) \widetilde{C}_{2}(q)= -\int\frac{\rmd{\bi q}}{(2\pi)^3}\widetilde{H}_{\rm A,2}(q) \widetilde{C}(q)
	\label{eq:lscatha2}
\end{equation}
where we used $\widetilde{C}_2(q)=\widetilde{P}_0(q)\widetilde{C}(q)$, see \eref{eq:lnprop} and \eref{eq:cnlm}, and introduced the extended $H_{\rm A}$-box
\begin{equation}
	\widetilde{H}_{\rm A,2}(q) = \widetilde{P}_{0}(q) \widetilde{H}_{\rm A}(q).
	\label{eq:ha2}
\end{equation}

\subsection{The transport mean free path}\label{sec:resultslt}
To calculate the corrections of the transport mean free path, we use \eref{eq:variancefullp}, which we rearrange in the form
\begin{equation}
	\frac{\ltrans}{\ell} = \frac{1}{2\lscat\ell} \int\rmd{\bi r}\, r^2 \Bigl( P_{0}(r) + P_{H}(r) + P_{B}(r) \Bigr). 
	\label{eq:rearrangedlt}
\end{equation}
Using \eref{eq:p0} we can evaluate the integral over the ladder step $P_{0}$ which yields
\begin{equation}
	\eqalign{
		\frac{\ltrans}{\ell} &= \left( \frac{\lscat}{\ell} \right)^{2} + \frac{1}{2\lscat\ell} \int\rmd{\bi r}\, r^2 \Bigl( P_{H}(r) + P_{B}(r) \Bigr) \\
		&\equiv \left( \frac{\lscat}{\ell} \right)^{2} + \frac{\delta H_{\rm t}}{\ell} + \frac{\delta B_{\rm t}}{\ell} \;,
	}
	\label{eq:ltransvslscat}
\end{equation}
where, analogously to the case of the scattering mean free path, we introduced the corrections due to the Hikami- and dressed ladder diagrams, $\delta H_{\rm t}$ and $\delta B_{\rm t}$, respectively. We can put \eref{eq:ltransvslscat} into a more convenient form by writing $\lscat=\ell + \delta\lscat$, and expand the square of $\lscat/\ell$ for small $\delta\lscat$ to obtain
\begin{equation}
	\eqalign{
		\frac{\ltrans}{\ell} &= -1 + 2 \frac{\lscat}{\ell} + \frac{\delta H_{\rm t}}{\ell} + \frac{\delta B_{\rm t}}{\ell}\\
			&= 1 + 2\frac{\delta H_{\rm s}}{\ell} + 2\frac{\delta B_{\rm s}}{\ell} + \frac{\delta H_{\rm t}}{\ell} + \frac{\delta B_{\rm t}}{\ell}.
	}
	\label{eq:convenient}
\end{equation}
We see that the corrections $\delta H_{\rm s}$ and $\delta B_{\rm s}$ of the scattering mean free path calculated in the previous section also enter into the corrections of the transport mean free path.

In order to use the exact ladder/crossed propagator in the calculation of the corrections, we again switch to the Fourier transforms and obtain:
\begin{equation}
	\eqalign{
		\frac{\delta H_{\rm t}}{\ell} = \frac{1}{2\ell^2} \int\frac{\rmd{\bi q}}{(2\pi)^3} H^{(2)}(q) \widetilde{C}(q) \\
		\frac{\delta B_{\rm A+\cdots+D,t}}{\ell} = \frac{1}{2\ell^2} \int\frac{\rmd{\bi q}}{(2\pi)^3} B^{(2)}(q) \widetilde{L}_{2}(q),
	}
	\label{eq:ltranscorrf}
\end{equation}
where we introduced the Fourier transforms $\widetilde{H}^{(2)}(q)$ and $\widetilde{B}^{(2)}(q)$ of the functions:
\begin{equation}
	H^{(2)}(r_{12}) = \int\rmd{\bi r}\rmd{\bi r}'\,\left| {\bi r} - {\bi r}' \right|^{2} H({\bi r}, {\bi r}_{1}, {\bi r}', {\bi r}_{2})
	\label{eq:fourierreducedweightedh}
\end{equation}
and
\begin{equation}
	B^{(2)}(r_{12}) = \int\rmd{\bi r}\rmd{\bi r}'\,\left| {\bi r} - {\bi r}' \right|^{2} B({\bi r}, {\bi r}_{1}, {\bi r}', {\bi r}_{2}),
	\label{eq:fourierreducedweightedl}
\end{equation}
respectively. As compared to $H(r_{12})$ and $B(r_{12})$ in \sref{s:lscat}, the additional term $\left| {\bi r} - {\bi r}' \right|^2$ appears in \eref{eq:fourierreducedweightedh} and \eref{eq:fourierreducedweightedl}. This makes the calculation more complicated, however -- with some effort -- the integrals in \eref{eq:fourierreducedweightedh} and \eref{eq:fourierreducedweightedl} can still be performed analytically.

We start with the contribution $H^{(2)}_{\rm A}$. Using translational invariance, we choose the point ${\bi r}_{2}$ in diagram \ref{fig:diagrams} a) as the origin. We then introduce the Fourier transforms of the Green's functions
\begin{equation}
	\eqalign{
		H_{\rm A}^{(2)}({\bi r}_{1}) =& \frac{4\pi}{\ell} \int\frac{\rmd{{\bi p}_{1}}\cdots\rmd{{\bi p}_{4}}}{(2\pi)^{12}}\rmd{{\bi r}}\rmd{{\bi r}'} \left| {\bi r}-{\bi r}' \right|^2 \\
		&\exp\bigl[\rmi{\bi p}_{1}\cdot{\bi r}-\rmi{\bi p}_{2}\cdot({\bi r}-{\bi r}_{1})+\rmi{\bi p}_{3}\cdot{\bi r}'-\rmi{\bi p}_{4}\cdot({\bi r}_{1}-{\bi r}')\bigr] \\
		&\widetilde{G}(p_1)\widetilde{G}^*(p_2)\widetilde{G}(p_3)\widetilde{G}^*(p_4).
	}
	\label{eq:har2-1}
\end{equation}
Substituting the variables ${\bi u} = ({\bi r} + {\bi r}')/2$ and ${\bi v} = {\bi r} - {\bi r}'$, shifting ${\bi p}_{2}$ by $-{\bi p}_{3}$, and using the relation
\begin{equation}
	\int\rmd{{\bi r}}\, r^2 \exp\bigr(\rmi{\bi p}\cdot{\bi r}\bigr) f({\bi p}) = -(2\pi)^3\delta({\bi p})\Delta_{\bi p}f({\bi p}),
	\label{eq:distr}
\end{equation}
valid for sufficiently smooth functions $f$, we arrive at the expression
\begin{equation}
	\eqalign{
		H_{\rm A}^{(2)}({\bi r}_{1}) =& -\frac{4\pi}{\ell} \int\frac{\rmd{{\bi p}_{3}}\rmd{{\bi p}_{4}}}{(2\pi)^6} \exp\bigl[-\rmi({\bi p}_{3}-{\bi p}_{4})\cdot{\bi r}_{1}\bigr] \\
		&\widetilde{G}(p_3)\widetilde{G}^*(p_4) \:\Delta_{{\bi p}_{2}}\left( \widetilde{G}({\bi p}_{2}+{\bi p}_{4})\widetilde{G}^*({\bi p}_{2}+{\bi p}_{3}) \right)\Bigr|_{{\bi p}_{2}=0} \;.
	}
	\label{eq:har2-2}
\end{equation}
The Laplacian of the product of Green's functions follows from \eref{eq:averagegreensfunctiontilde}:
\begin{equation}
	\eqalign{
	\fl\Delta_{{\bi p}_{2}}\left( \widetilde{G}({\bi p}_{2}+{\bi p}_{4})\widetilde{G}^*({\bi p}_{2}+{\bi p}_{3}) \right)\Bigr|_{{\bi p}_{2}=0}
		&= 6 \widetilde{G}^2(p_4)\widetilde{G}^*(p_3) + 6 \widetilde{G}(p_4)\widetilde{G}^{*2}(p_4) \\
		&+ 8 p_4^2 \widetilde{G}^3(p_4)\widetilde{G}^*(p_3)+ 8 p_3^2 \widetilde{G}(p_4)\widetilde{G}^{*3}(p_3)\\
	&+ 8 {\bi p}_{4}\cdot{\bi p}_{3} \widetilde{G}^2(p_4)\widetilde{G}^{*2}(p_3).
	}
	\label{eq:har2-3}
\end{equation}
Choosing spherical coordinates, one can first perform the integration over the angular variables, and then evaluate the remaining integrals over $p_{3}$ and $p_{4}$ using the calculus of residues. Finally, Fourier transforming the result with respect to ${\bi r}_{1}$ yields the desired $\widetilde{H}^{(2)}_{\rm A}(q)$. From this, the correction is obtained by numerical integration of $\widetilde{H}^{(2)}_{\rm A}(q) \widetilde{C}(q)$ over ${\bi q}$ (cf.\ \eref{eq:ltranscorrf}).

Like in \sref{s:lscat}, the short loop (diagram in \fref{fig:leadingdiagram}) accounts for the leading order correction and can be obtained analytically again by replacing the full crossed propagator $C$ with a single step $4\pi P_{0}/\ell$:
\begin{equation}
	\eqalign{
		\frac{\delta H_{\rm A,1,t}}{\ell} &= \frac{1}{2\ell^2} \int\frac{\rmd{\bi q}}{\left(2\pi\right)^{3}} \widetilde{H}^{(2)}_{\rm A}(q) \frac{4\pi}{\ell}\widetilde{P}_{0}(q) \\
		&= \frac{1}{24 k^4\ell^3\lscat} \Biggl[ \frac{(k\lscat)^2 + 4 (k\lscat)^4}{1 + (k\lscat)^2} + \left( -6 k\lscat + 40 (k\lscat)^3 \right) \arctan(k\lscat)\\
			&\phantom{=\,} + \left( 5 - 12 (k\lscat)^2 \right) \ln\bigl(1 + (k\lscat)^2\bigr) \Biggr] \underset{k\ell\gg 1}{\longrightarrow}\frac{5\pi}{6k\ell} - \frac{\ln(k\ell)}{(k\ell)^2}.
	}
	\label{eq:ltrha1}
\end{equation}
The remaining contribution to $\delta H_{\rm A,t}$, i.e.\ $\delta H_{\rm A,2,t}\equiv \delta H_{\rm A,t}-\delta H_{\rm A,1,t}$, reads:
\begin{equation}
		\frac{\delta H_{\rm A,2,t}}{\ell} = \frac{1}{2\ell^2} \int\frac{\rmd{{\bi q}}}{(2\pi)^3} \widetilde{H}_{\rm A}^{(2)}(q) \widetilde{C}_{2}(q) = \frac{1}{2\ell^2} \int\frac{\rmd{{\bi q}}}{\left(2\pi\right)^{3}} \widetilde{H}_{\rm A,2}^{(2)}(q) \widetilde{C}(q) ,
	\label{eq:ha2fullpl}
\end{equation}
again with the extended $H_{\rm A,2}^{(2)}$-box given by, cf.\ \eref{eq:ha2},
\begin{equation}
	\widetilde{H}_{\rm A,2}^{(2)}(q) = \widetilde{P}_{0}(q) \widetilde{H}_{\rm A}^{(2)}(q).
	\label{eq:ha22}
\end{equation}

Like $\delta H_{\rm A,1,t}$, the contribution $\delta B_{\rm E+F,t}$ from diagrams \ref{fig:diagrams} h) and i) can also be calculated analytically:
\begin{equation}\label{eq:deltabeft}
\eqalign{
    \fl\frac{\delta B_{\rm E+F,t}}{\ell} = \frac{1}{24 (k\ell)^4 \bigl[1 + 5 (k\ell)^2 + 4 (k\ell)^4\bigr]} \biggl\{ 2 (k\ell)^2 + 52 (k\ell)^4 + 32 (k\ell)^6 \\
        + 4 k\ell \left[27 + 115 (k\ell)^2 + 8 (k\ell)^4 - 80 (k\ell)^6\right] \arctan(k\ell) \\
        - 20 k\ell \left[3 + 11 (k\ell)^2 - 8 (k\ell)^4 - 16 (k\ell)^6\right] \arctan(2 k\ell) \\
        - \left(10 - 46 (k\ell)^2 - 440 (k\ell)^4 - 384 (k\ell)^6\right) \ln\bigl(1 + (k\ell)^2\bigr) \\
        + \left(5  - 35 (k\ell)^2 - 280 (k\ell)^4 - 240 (k\ell)^6\right) \ln\bigl(1 + 4 (k\ell)^2\bigr) \biggr\}\\
        \underset{k\ell\gg 1}{\longrightarrow} \frac{3 \ln\left( k\ell \right)}{(k\ell)^2} + \frac{2 - 5 \ln\left( 2 \right)}{(k\ell)^2}\;.
}
\end{equation}

\begin{figure}
	\begin{indented}
		\item[]\input{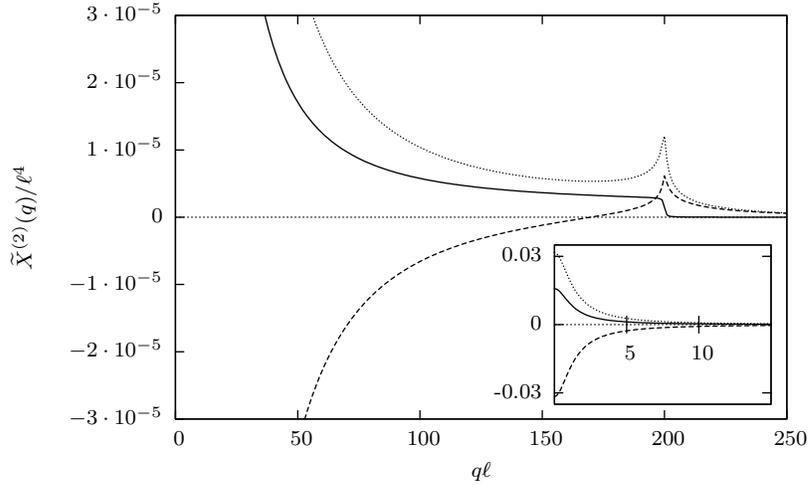}
	\end{indented}
    \caption{The functions $\widetilde{H}^{(2)}_{\rm A,2}(q)$ (solid), $\widetilde{H}^{(2)}_{\rm B+C}(q)=\widetilde{B}^{(2)}_{\rm A+B}(q)$ (dashed), and $\widetilde{B}^{(2)}_{\rm C+D}(q)$ (dotted), from the end of \sref{sec:resultslt}, for the disorder parameter $k\ell=100$. Due to the very lengthy nature of the analytical expressions of these functions they are not given explicitely in the text. The inset shows that the functions remain finite for vanishing $q$.}
	\label{fig:functions}
\end{figure}

For the contributions from the remaining diagrams, \ref{fig:diagrams} b)--g), one can proceed analogously to the calculations \eref{eq:har2-1}-\eref{eq:har2-3}. Since the resulting expressions are very lengthy, we do not give them explicitely but rather discuss their general behaviour. A plot of the Fourier transforms of the functions $\widetilde{H}^{(2)}_{\rm A,2}$, $\widetilde{H}^{(2)}_{\rm B+C}=\widetilde{B}^{(2)}_{\rm A+B}$ and $\widetilde{B}^{(2)}_{\rm C+D}$, as functions of momentum $q$, is given in \fref{fig:functions}. These functions are directly comparable with each other, since their contributions to the correction of the transport mean free path are given by integrating these functions multiplied by $\widetilde{C}(q)=\widetilde{L}_2(q)$ over the momentum $q$, see \eref{eq:ltranscorrf} and \eref{eq:ha2fullpl}. The inset of \fref{fig:functions} shows that all functions remain finite for vanishing $q$. For these limiting values we obtain, in leading order of $1/k\ell$:
\numparts
\begin{eqnarray}
    \widetilde{H}^{(2)}_{\rm A,2}(0) &\approx& \left( \frac{4\pi\ell}{k} \right)^2\\
    \widetilde{H}^{(2)}_{\rm B+C}(0) = \widetilde{B}^{(2)}_{\rm A+B}(0) &\approx -2& \left( \frac{4\pi\ell}{k} \right)^2\\
    \widetilde{B}^{(2)}_{\rm C+D}(0) &\approx \phantom{-}2&\left( \frac{4\pi\ell}{k} \right)^2\;.
\end{eqnarray}
\endnumparts
From \fref{fig:functions}, we also see that the function $\widetilde{H}^{(2)}_{\rm A,2}(q)$ is cut off at $q\approx 2k$, whereas the functions $\widetilde{H}^{(2)}_{\rm B+C}(q)$, $\widetilde{B}^{(2)}_{\rm A+B}(q)$ and $\widetilde{B}^{(2)}_{\rm C+D}(q)$ are peaked at $q\ell=\sqrt{1+(2k\ell)^2}\underset{k\ell\gg 1}{\approx} 2k\ell$, and then decay like $1/q$, asymptotically.

In order to verify the consistency of our results, the values at $q=0$ given above can also be derived from the ordinary boxes $\widetilde{H}_{\rm A,B,C}$ and $\widetilde{B}_{\rm A,\dots,D}$, see \eref{eq:haq}, \eref{eq:hbqlaq}, and \eref{eq:lcqldq}, by using the following identity:
\begin{equation}
	\mathcal{F}\left\{ r^2 f(r) \right\}(q)\Big\vert_{q=0} = -\Delta_{\bi q} \widetilde{f}(q)\Big\vert_{q=0} = -2 \frac{\widetilde{f}'(q)}{q}\Big\vert_{q=0} - \widetilde{f}''(q)\Big\vert_{q=0} \;,
	\label{eq:check}
\end{equation}
valid for the Fourier transformation $\mathcal{F}$ of a spherically symmetric, twice differentiable function $f$ multiplied by $r^2$. This relation can be derived with the help of \eref{eq:distr}. 

\section{Results}\label{s:results}

In the previous section, we have derived expressions for the different contributions to the corrections of the scattering and the transport mean free path. Except for the cases of the special diagrams shown in figures \ref{fig:leadingdiagram}, \ref{fig:diagrams} h) and i), which are given analytically, the integrals in these expressions have not yet been evaluated, cf.\ \eref{eq:deltah}, \eref{eq:deltal}, \eref{eq:lscatha2}, \eref{eq:ltranscorrf} and \eref{eq:ha2fullpl}. We now analyse the asymptotic behaviour of these contributions, by considering the data obtained from numerical integration.

\subsection{The scattering mean free path}\label{ss:lscat}
The leading asymptotic behaviour of \eref{eq:lscatha1}, i.e.\ the correction due to the diagram depicted in \fref{fig:leadingdiagram}, is $\delta H_{\rm A,1,s}/\ell \longrightarrow -\pi/2k\ell$ for $k\ell\rightarrow\infty$. In higher order, it also contributes corrections of the order of $\ln\left( k\ell \right)/\left( k\ell \right)^2$, $1/\left( k\ell \right)^2$, and further.

The remaining diagrams give corrections at least of the order of $\ln\left( k\ell \right)/\left( k\ell \right)^2$ and $1/\left( k\ell \right)^2$ or higher. We thus write:
\begin{eqnarray}
	\frac{\delta H_{\rm s}}{\ell},\frac{\delta B_{\rm s}}{\ell} \longrightarrow \frac{a_{1} \ln(k\ell) + a_{2}}{(k\ell)^2} &\qquad& \mathrm{for\ } k\ell\rightarrow\infty.
	\label{eq:lscatasymptotic}
\end{eqnarray}
\Tref{tab:coefflscat} lists the coefficients $a_{1}$ and $a_{2}$ in \eref{eq:lscatasymptotic} for the individual contributions obtained analytically from \eref{eq:corrlscatlefasymptotic} and \eref{eq:lscatha1} (for $\delta B_{\rm E+F,s}$) or by fitting the numerically integrated contributions (cf.\ \eref{eq:deltah} and \eref{eq:deltal}). We see that the different logarithmic contributions cancel each other (cf.\ the leading coefficient $a_{1}$ in \tref{tab:coefflscat}), which only leaves contributions of the order of $1/(k\ell)^{2}$. However, due to the contribution of $\delta H_{\rm A,1,s}$ \eref{eq:lscatha1}, a term of the order of $1/k\ell$, which thus constitutes the leading contribution, remains. The total correction of the scattering mean free path then reads
\begin{eqnarray}
    \frac{\delta \lscat}{\ell} \longrightarrow -\frac{\pi}{2k\ell} + \Or((k\ell)^{-2}) &\qquad&\phantom{mmm} \mathrm{for\ } k\ell\rightarrow\infty.
	\label{eq:ha1lscatasymptotic}
\end{eqnarray}
This is the same correction one would find with the help of \eref{eq:deflscat} from the second term of the self-energy depicted in \fref{fig:sigmau}, see e.g.\ chapter 3.2 exercise 3.8 equation (3.82) in \cite{akkermans:mesoscopic}. As already mentioned, this leading contribution originates from the short-loop diagram shown in \fref{fig:leadingdiagram}, where only a single step $4\pi \widetilde{P}_0/\ell$ is inserted instead of the full propagator $\widetilde{C}$ in the Hikami box $\widetilde{H}_{\rm A}$, see \eref{eq:lscatha1}. As we have checked, inserting two steps $4\pi\widetilde{P}_0^2/\ell$ into $\widetilde{H}_{\rm A}$ -- or, equivalently, a single step $4\pi\widetilde{P}_0/\ell$ into the extended box $\widetilde{H}_{\rm A,2}$, see \eref{eq:lscatha2} -- reproduces the next order correction, i.e., the coefficient $a_1=-1.234$ of $\delta H_{\rm A,2,s}/\ell$. Likewise, the $a_1$-coefficients of $\delta H_{\rm B+C,s}/\ell$, $\delta B_{\rm A+B,s}/\ell$ and $\delta B_{\rm C+D,s}/\ell$ (which, as mentioned above, cancel each other together with $\delta H_{\rm A,2,s}/\ell$), are obtained by inserting a single step into the boxes $\widetilde{H}_{\rm B+C}$, $\widetilde{B}_{\rm A+B}$, and $\widetilde{B}_{\rm C+D}$, respectively, i.e., \eref{eq:deltah} and \eref{eq:deltal} with $4\pi\widetilde{P}_0/\ell$ instead of $\widetilde{C}=\widetilde{L}_2$. Therefore, all contributions from longer loops between $\bi r_1$ and $\bi r_2$ (see \fref{fig:diagrams}), i.e., those obtained by inserting a full propagator $\widetilde{C}_2=\widetilde{L}_3$ with at least three steps into the boxes $\widetilde{H}_{\rm A,2}$, $\widetilde{H}_{\rm B+C}$, $\widetilde{B}_{\rm A+B}$, and $\widetilde{B}_{\rm C+D}$, yield corrections of the order $1/(k\ell)^2$. As we have checked, the $a_2$-coefficients of these longer loops also cancel each other. The same remains true if these longer loops are approximated by a diffusion propagator (proportional to $1/q^2$). For this reason, no correction of the scattering mean free path is found if the return probability is evaluated within the diffusion approximation, as mentioned in the introduction. In contrast, our exact calculation indicates a non vanishing correction of $\lscat$ due to short loops. Let us note that, in a similar context, the influence of short scattering paths on weak localisation in quantum billiards has recently been analysed in \cite{brezinova:prb77,brezinova:prb81}. 

\begin{table}
    \caption{Coefficients $a_{1}$ and $a_{2}$ defining the corrections of the scattering mean free path in orders $\ln(k\ell)/(k\ell)^2$ and $1/(k\ell)^2$, see \eref{eq:lscatasymptotic}, for the different contributions from the diagrams in \fref{fig:diagrams}. The data are obtained either analytically from \eref{eq:corrlscatlefasymptotic} and \eref{eq:lscatha1} or by fitting from \eref{eq:deltah}, \eref{eq:deltal} and \eref{eq:lscatha2}.}
\label{tab:coefflscat}
	\begin{indented}
	\item[]\begin{tabular}{llllll}
		\br
						& $\delta H_{\rm A,1,s}/\ell$ & $\delta H_{\rm A,2,s}/\ell$	&	$\delta H_{\rm B+C,s}/\ell$	& $\delta B_{\rm C+D,s}/\ell$		&	$\delta B_{\rm E+F,s}/\ell$ \\
						&						&											&	$= \delta B_{\rm A+B,s}/\ell$	&																	&	\\\mr
		$a_{1}$	&	$1$	& $-1.234$	&	$1.234$	&	$-1.234$		&	$-1$	\\
		$a_{2}$	&	$1+\pi^2/2$	&	$-1.206$	&	$0.155$	&	$-2.258$	&	$-1+\ln(2)$	\\
		\br
	\end{tabular}
	\end{indented}
\end{table}

\subsection{The transport mean free path}\label{ss:ltr}
As in the case of the scattering mean free path, the leading contribution originates from the diagram in \fref{fig:leadingdiagram} which is given by \eref{eq:ltrha1}. The leading asymptotic behaviour of this contribution is $\delta H_{\rm A,1,t}/\ell \longrightarrow 5\pi/6k\ell$, for $k\ell\rightarrow\infty$. Like in the case of the scattering mean free path, this diagram also contributes to the logarithmic term ($\ln(k\ell)/(k\ell)^2$) and terms of order $1/(k\ell)^2$ and higher.

The remaining contributions are at least of logarithmic form and we thus write their asymptotic behaviour like in \eref{eq:lscatasymptotic}:
\begin{eqnarray}
	\frac{\delta H_{\rm t}}{\ell},\frac{\delta B_{\rm t}}{\ell} \longrightarrow \frac{b_{1} \ln(k \ell) + b_{2}}{(k\ell)^2} &\qquad& \mathrm{for\ } k\ell\rightarrow\infty.
	\label{eq:ltransasymptotics}
\end{eqnarray}
\Tref{tab:coeffltrans} lists the coefficients $b_{1}$ and $b_{2}$ obtained either analytically (for $\delta B_{\rm E+F,t}$) or by fitting \eref{eq:ltransasymptotics} on the numerically integrated contributions (cf.\ \eref{eq:ltranscorrf} and \eref{eq:ha2fullpl}).
\begin{table}
    \caption{Coefficients $b_{1}$ and $b_{2}$ defining the corrections of the transport mean free path in orders $\ln(k\ell)/(k\ell)^2$ and $1/(k\ell)^2$, see \eref{eq:ltransasymptotics}, for the different contributions from the diagrams in \fref{fig:diagrams}. The data are obtained either analytically from \eref{eq:ltrha1} and \eref{eq:deltabeft} or by fitting from \eref{eq:ltranscorrf} or \eref{eq:ha2fullpl}, respectively.}
\label{tab:coeffltrans}
	\begin{indented}
	\item[]\begin{tabular}{llllll}
		\br
				& $\delta H_{\rm A,1,t}/\ell$		& $\delta H_{\rm A,2,t}/\ell$	&	$\delta H_{\rm B+C,t}/\ell$	& $\delta B_{\rm C+D,t}/\ell$		&	$\delta B_{\rm E+F,t}/\ell$ \\
						&		&																	&	$= \delta B_{\rm A+B,t}/\ell$	&																	&	\\\mr
		$b_{1}$	&	$-2$	&	$1.234$	&	$-2.968$	&	$2.967$		&	$3$	\\
		$b_{2}$	&	$-(3/2+5\pi^2/4$)	&	$2.439$	&	$0.906$	&	$4.537$		&	$2-5\ln(2)$	\\
		\br
	\end{tabular}
	\end{indented}
\end{table}

Using \eref{eq:convenient} and collecting all terms up to order $\ln(k\ell)/(k\ell)^2$ we obtain the correction of the transport mean free path
\begin{eqnarray}
    \frac{\delta\ltrans}{\ell} \longrightarrow -\frac{\pi}{6k\ell} - \frac{0.734 \ln(k\ell)}{(k\ell)^2} + \Or((k\ell)^{-2})&\qquad& \mathrm{for\ } k\ell\rightarrow\infty.
	\label{eq:fulldeltaltrans}
\end{eqnarray}
Similar to the case of the scattering mean free path (cf.\ end of \sref{ss:lscat}), we analysed the behaviour of the weak localisation corrections due to long loops with at least three scattering events. This is achieved by inserting $\widetilde{C}_{2}=\widetilde{L}_{3}$, instead of the full $\widetilde{C}$, into the extended box $H_{\rm A,2}^{(2)}$ or the other boxes $H_{\rm B+C}^{(2)}$, $B_{\rm A+B}^{(2)}$ and $B_{\rm C+D}^{(2)}$, and leads to vanishing logarithmic terms $b_{1}$. As for the remaining coefficient $b_{2}$, the contributions of the ladder boxes $B_{\rm A+B}^{(2)}$ and $B_{\rm C+D}^{(2)}$ cancel each other. In contrast, the Hikami boxes $H_{\rm A,2}^{(2)}$ and $H_{\rm B+C}^{(2)}$ do not compensate for each other and the resulting correction of the transport mean free path in this case reads $\delta\ltrans/\ell\approx-4.4/(k\ell)^2$. This shows that the long loops contribute only to order $1/(k\ell)^2$, which is consistent with the standard result obtained within the diffusion approximation, see e.g.\ \cite{akkermans:mesoscopic} chapter 7.4 (7.57).

A comparison of the $1/k\ell$ contribution in \eref{eq:fulldeltaltrans} with \eref{eq:ha1lscatasymptotic} shows that $\lscat$ is more strongly reduced than $\ltrans$. The relative increase of the transport mean free path can be made plausible by looking more closely at the leading diagram, see \fref{fig:leadingdiagram}, and its effect on the step length distribution \eref{eq:steplength}. It turns out that the main contribution to this diagram comes from constellations where ${\bi r}_{1}$ and ${\bi r}_{2}$ are very close to each other. This approximately amounts to two independent steps, each of which follows the exponential step length distribution $P_{0}$ (cf.\ \eref{eq:p0}), at once. Due to this ``double step'' the variance of the step length distribution, and thus the transport mean free path, is increased with respect to the case with just single steps.

We finally look at the weak localisation corrections of the conductivity which were calculated in \cite{wysokinski:pre52}:
\begin{equation}
	\eqalign{
    \sigma &= \sigma_{0} \left( 1 - \frac{2\pi}{3k\ell} - \frac{\pi^2-4}{8}\frac{\ln(k\ell)}{\left( k\ell \right)^2} + \Or((k\ell)^{-2})\right) \\
			&\approx \sigma_{0} \left( 1 - \frac{2\pi}{3k\ell} - \frac{0.734 \ln(k\ell)}{\left( k\ell \right)^2} + \Or((k\ell)^{-2}) \right).
	}
	\label{eq:sigmabelitz}
\end{equation}
We compare this with the quantity $\mathcal{D}^{\rm (s)}=\lscat\ltrans/3$ for which we find, putting all previous corrections of $\lscat$ and $\ltrans$ together, the following leading order weak localisation corrections:
\begin{equation}
	\mathcal{D}^{\rm (s)} = \frac{\ell^2}{3}\left( 1 - \frac{2\pi}{3k\ell} - \frac{0.734 \ln(k\ell)}{\left( k\ell \right)^2} + \Or((k\ell)^{-2}) \right).
	\label{eq:belitz}
\end{equation}
Hence, the leading weak disorder corrections of the conductivity $\sigma$ and of the stationary diffusion constant $\mathcal{D}^{\rm (s)}$ are identical. Since, to the best of our knowledge, a general relation between $\mathcal{D}^{\rm (s)}$ and $\sigma$ has not yet been proven, it is an interesting open question whether the proportionality between $\mathcal{D}^{\rm (s)}$ and $\sigma$ remains valid also for stronger disorder (i.e. beyond the leading orders in $1/k\ell$).

\section{Conclusion \& Outlook}
We calculated the leading weak disorder corrections of the scattering and of the transport mean free path for the propagation of waves in a white-noise random potential, without employing any approximation except $k\ell \gg 1$ (where $\ell$ is the Boltzmann mean free path). The leading corrections for both, $\lscat$ and $\ltrans$, scale like $1/(k\ell)$, and originate from the short-loop diagrams given in \fref{fig:leadingdiagram}. In contrast, the self-consistent theory of localisation \cite{vollhardt:prb22,tiggelen:prl84}, where the loops are treated within the diffusion approximation, predicts a vanishing correction of $\lscat$, whereas the correction of $\ltrans$ scales like $1/(k\ell)^2$. This is not necessarily a contradiction, since the self-consistent theory of localisation is meant to describe the transition from the regime of weak to strong localisation, rather than to calculate corrections for the case of weak disorder. The basic idea of the self-consistent theory is the renormalisation of the diffusion constant, where the loops contributing to the return probability (which, in turn, determines the diffusion constant) are treated in the diffusion approximation, but with a renormalized diffusion constant. Since the present paper shows that also short loops give rise to significant weak disorder corrections for both, $\lscat$ and $\ltrans$, this may motivate future attempts to improve the self-consistent theory of localisation (which, in its present form, does not reproduce the correct critical exponents for the Anderson transition \cite{vollhardt:review1992,schreiber:prl76}), e.g.\ by renormalizing not only a single quantity describing transport on large length scales (such as the diffusion constant or the transport mean free path), but taking into account also quantities relevant for transport on shorter length scales (such as the scattering mean free path). 

Furthermore, let us comment on the relevance of our results with respect to experiments on weak localisation \cite{bergmann:physrep107,adams:prb45}. The experiments described in \cite{bergmann:physrep107} consider the weak localisation corrections of the electron conductivity (or, similarly, the mobility) under breaking of the reciprocity symmetry, whereas the disorder parameter $1/k\ell$ is kept constant. Breaking of the reciprocity symmetry is achieved by the application of a magnetic field which affects mainly the long interference loops, since these encircle on average large areas being sensitive to penetration by the magnetic flux. Thus, these experiments observe the weak localisation contribution of order $1/(k\ell)^2$ characteristic for long loops (the logarithmic correction being hard to distinguish from the pure $1/(k\ell)^2$ decay, experimentally). However, we showed that the diagram depicted in \fref{fig:leadingdiagram} gives rise to a correction of the order $1/k\ell$. Again, this is no contradiction since this diagram is short range and thus almost insensitive to reciprocity breaking induced by the magnetic field. Indeed, other experiments \cite{schwarz:prb21}, which measure the mobility of electrons in helium gas and directly change the density of the scattering medium and thus the disorder parameter $1/k\ell$, were confirmed to be consistent with a leading order correction to scale like $1/k\ell$ in \cite{adams:prb45}.

Based on the thorough understanding of the linear case, as provided in the present article, it becomes now furthermore possible to investigate the influence of \emph{non-linearities} on the weak localisation effect. Application of the non-linear diagrammatic theory developed in \cite{wellens:prl100,wellens:pra80} to the case of non-linear coherent backscattering showed an inversion of the CBS cone, already for moderate values of the non-linearity \cite{wellens:nlcbs}. This inversion stems from \emph{destructive} interference in the backscattering direction and thus suggests a weak anti-localisation effect. This cone inversion was also observed in numerical experiments on coherent backscattering of Bose-Einstein condensates in two dimensional speckle potentials \cite{hartung:prl101}, in accordance with the non-linear diagrammatic theory. Combining the non-linear diagrammatic theory and the work presented in this article it is possible to calculate the corrections of the transport mean free path in the non-linear case. The negative cone should then correspond to an increased transport mean free path as compared to the classical (ladder approximation) prediction.

\ack
We would like to thank Nicolas Cherroret and Cord A.\ M\"uller for fruitful discussions. We also acknowledge financial support by the ``Deutsche Forschungsgemeinschaft'' under the grant number \emph{DFG BU1337/8-1} and the ``DFG Forschergruppe 760: Scattering Systems with Complex Dynamics''.

\bibliographystyle{iopart-num}
\bibliography{Bibliography-shortpaths}

\end{document}